\documentclass[namedreferences]{solarphysics}

\usepackage[hyperref,optionalrh,showbiblabels]{spr-sola-addons} 
\usepackage{graphicx}        
\usepackage{color}           
\usepackage{breakurl}        

\chardef\us=`\_
\begin{document}

\begin{article}
\begin{opening}

\title{Harmonics of Solar Radio Spikes at Metric Wavelengths}

\author[addressref={aff1,aff2},corref,email={winfeng@sdu.edu.cn}]{\inits{S. W.}\fnm{S. W.}~\lnm{Feng}\orcid{https://orcid.org/0000-0002-9634-5139}}

\author[addressref=aff1]{\inits{Y.}\fnm{Y.}~\lnm{Chen}\orcid{ https://orcid.org/0000-0001-6449-8838}}
\author[addressref=aff1]{\fnm{C. Y.}~\lnm{Li}}
\author[addressref=aff1]{\fnm{B.}~\lnm{Wang}}
\author[addressref=aff1]{\fnm{Z.}~\lnm{Wu}}
\author[addressref=aff1]{\fnm{X. L.}~\lnm{Kong}\orcid{https://orcid.org/0000-0003-1034-5857}}
\author[addressref={aff1,aff3}]{\fnm{Q. F.}~\lnm{Du}}
\author[addressref={aff1,aff3}]{\fnm{J. R.}~\lnm{Zhang}}
\author[addressref=aff4]{\fnm{G. Q.}~\lnm{Zhao}\orcid{https://orcid.org/0000-0002-1831-1451}}


\address[id=aff1]{Shandong Provincial Key Laboratory of Optical Astronomy and
Solar-Terrestrial Environment, and Institute of Space Sciences, Shandong University, Weihai 264209, China}
\address[id=aff2]{State Key Laboratory of Space Weather, Chinese Academy of Sciences, Beijing 100190, China}
\address[id=aff3]{School of Mechanical, Electrical \& Information Engineering, Shandong University, Weihai 264209, China}
\address[id=aff4]{Institute of Space Physics, Luoyang Normal University, Luoyang 471022, China}

\runningauthor{S. W.\ Feng \emph{et al.}}
\runningtitle{Solar Radio Spike Harmonics}

\begin{abstract}
This paper presents the latest observations from the newly-built
solar radio spectrograph at the \emph{Chashan
Solar Observatory}. On July 18 2016, the spectrograph records
a solar spike burst event, which has several episodes showing
harmonic structures, with the second, third, and fourth harmonics.
The lower harmonic radio spike emissions are observed later than
the higher harmonic bands, and the temporal delay of the second
(third) harmonic relative to the fourth harmonic is about 30\ --\ 40 (10) ms.
Based on the electron cyclotron maser emission mechanism, we analyze
possible causes of the temporal delay and further infer relevant coronal
parameters, such as the magnetic field strength and the electron
density at the radio source.
\end{abstract}
\end{opening}
\section{Introduction}
     \label{S-Introduction}
On the solar radio dynamic spectrum, spikes are narrowband bursts
with typical relative bandwidth being about 1\ --\ 3 \% and  duration $<$ 100 ms.
Major observational features are as follows: narrowband type-III like storms,
spikes, dots, sub-second clusters, groups and chains, and other narrowband
structures (\citealp{Benz1986}; \citealp{Bouratzis2015,Bouratzis2016}). Radio
spikes often present as fine structures in the continuum of type IV bursts.

Radio spikes are closely related to the process of energy release in solar
flares, and can be used to infer coronal parameters and the energy release
process such as magnetic reconnection. Since the discovery of radio spikes
in 1961, they have been  extensively studied from decimeter to decameter wavelengths ({\it e.g.} \citealp{Benz1986};
\citealp{fleishman1998}; \citealp{chernov2011}; \citealp{Melnik2014};
 \citealp{Bouratzis2015,Bouratzis2016}; \citealp{Shevchuk2016}).

At metric wavelengths the typical duration of radio spikes is 10\ --\ 100 ms,
the  bandwidth is 0.5\ --\ 15 MHz (with a relative bandwidth being $\approx$ 2 \%),
the degree of circular polarization is high,
and the spectral drift rate can be either positive, negative, or none. Recently,
\cite{Bouratzis2016}  studied 12000 metric radio spikes using observational
data from ARTEMIS-Jean Louis Steinberg Radiospectrograph (ARTEMIS-IV) with the temporal resolution of 10 ms and the frequency
range of 270\ --\ 450 MHz, and found that the duration of  these events was 60 ms
and the average relative bandwidth was 2 \%. They also presented the U-shaped,
J-shaped, point-like, and cluster-chain spikes.

Harmonic structures of spikes were reported for
the first time by \cite{Benz1987}. They studied 36 events of radio spikes
using observational data of the Zurich analog spectrograph  with the frequency
range of 100\ --\ 1000 MHz and the time resolution of 25\ --\ 100 ms, and found
that for one event there are two emission bands at different
frequencies having similar morphology. The frequency ratio of the two
emission bands is 1:1.39. It is suggested that this two emission
bands correspond to different harmonics. Later,
\cite{gudel1990} and \cite{Krucker1994} also studied the harmonic characteristics
of spikes. \cite{gudel1990} analyzed the data observed by two frequency-agile
receivers near Zurich with the frequency range of 100\ --\ 3000 MHz and a temporal cadence
of 50\ --\ 250 ms. In the paper, the author gave an example of harmonic structures
in nine episodes of solar radio spikes between 1980 and 1990. It is
found that the harmonics are not limited to the second and third orders,
the fifth, sixth, or even eighth harmonic orders also exist. The
fundamental (not detected) frequency of these events lies in the
range of 170\ --\ 350 MHz. The intensity of the radio spikes becomes stronger
with the increase of harmonic orders. The author also attempted to measure the time
delay of lower harmonics relative to higher ones, and found that the
delay was likely less than 50 ms. That is below the temporal resolution
of the observational data. Therefore, the exact delay of the lower-relative-to
higher harmonics could not be determined.

The ARTEMIS-IV data have a higher temporal
resolution (10 ms). Yet, its observing band is from 270 to 450 MHz, too
narrow to observe the full harmonics of
radio spikes. This should be the reason why \cite{Bouratzis2015,Bouratzis2016}
did not observe any harmonic structures of radio spikes.

The temporal delay of the lower-relative-to higher harmonics of radio
spikes is yet to be determined. So far, the observational data used in the
study of radio spike harmonic structures are observed in the 1970s
and 1980s. The frequency range and the time resolution of those data are
somewhat insufficient to measure the relative timings of various harmonic structures.

In this paper, we use the latest data from a new solar radio dynamic
spectrograph working at metric wavelengths to study solar radio spike
harmonic structures. The following section is about the instrument
and data. In the third section, we present the observational
characteristics of the event. In the fourth section, we discuss possible
causes of the time delay of lower-order harmonics. A summary is presented
in the last section.

\section{Observational Data and Instrument}
      \label{S-general}
In this paper, we mainly use the data from the solar radio dynamic
spectrograph at the \emph{Chashan Solar Observatory} (CSO).
The station is located at $122.3^{\circ}$E and $36.8^{\circ}$N, and is
constructed and managed by the \emph{Institute of Space Sciences of Shandong University}.
On its northern side, there is the \emph{Chashan} mountain extending along the east-west direction of
more than 500 meters high above the sea level, and on the southern side it faces the
 \emph{Yellow Ocean}. The mountain blocks a significant part of broadcast communication
 signals and other radio interferences from the city. Therefore, the radio
frequency interference is at a relatively low level. This makes it a nice station to
observe solar radio bursts at metric wavelengths.

The newly built solar radio observational system includes a 6-meter parabolic
dish fed by a dual-polarized log-periodic antenna as shown in Figure 1. The
observing band is 150\ --\ 500 MHz and the beam size is  12$^{\circ}$ at 300 MHz. The
received signals are filtered, amplified, and sampled by an AD converter, and
then FFT analysis is performed with a high-speed Field Programmable Gate Array (FPGA).
The resolutions of the obtained radio dynamic
spectral data are 10 ms in time and 160 kHz in frequency, and the dynamic range is $\approx$ 50 dB.
\cite{du2017} gives the technical details of the data acquisition and recording. In order to make sure the
observational data are reliable, we also use the data of the ORFEES
(\url{http://radio-monitoring.obspm.fr/downloadOrfeesHR1.php}) and the
\emph{Radio Solar Telescope Network} (RSTN)/Learmoth
(\url{http: //www.sws. Bom.gov.au/World_Data_Centre/1/10}) for comparison.

The solar radio burst occurred on July 18 2016. The associated active
region is located at N05W03, and the soft X-ray flare class is C4.4 with
the start, peak, and end time of 08:09:00 UT, 08:23:00 UT, and 08:32:00 UT.
Radio spikes studied in this article were observed during
08:10:00 UT\ --\ 08:15:00 UT, in the pre-impulsive phase of the flare.
Figures 2a and 2b show the solar radio dynamic spectra of these
spikes recorded by both ORFEES and CSO. For comparison, ORFEES and CSO data are adjusted
to the same time resolution and frequency range. Obviously, the radio
dynamic spectra from this two stations are basically the same.
The spectra contain lots of narrow band spikes superposed onto a
type IV continuum burst.

To further confirm the accuracy of the observed data, we compared the
CSO data with the calibrated RSTN/Learmoth (LEAR) data. RSNT/LEAR
can provide solar observation data at 10 frequencies between 245\ --\ 15000 MHz, and only the
two frequencies of 245 MHz and 410 MHz  lie in the CSO
observational band. The flux density profiles from LEAR
at these two frequencies are shown in Figures 2c\ --\ d as
solid lines. The recorded signal values $R$ of CSO at the two corresponding
frequencies are adjusted according to the relationship of
\begin{equation}
R'=A\log[(R-B)C],
\end{equation}
where $A$ and $C$ are mainly determined by the receiver system gain, and $B$ is mainly
determined by the receiver system noise. A similar calibration method has been
used by \cite{messmer99}.  We adjust the values of $A$, $B$,
and $C$  so as to make the CSO flux density profiles as close as possible to the LEAR profiles. The values $R'$
are shown by dotted lines at the corresponding frequencies in
Figures 2c\ --\ d. It is found that we can make the light curves of CSO
data basically the same as those of LEAR at the two frequencies.
This can be regarded as a preliminary calibration process. Further
calibration of the system with calibrated noise sources and signals is under way.

\section{Metric Radio Spikes Observed by CSO}
Figure 3 shows a segment of the spectrum with the highest temporal
resolution of 10 ms observed by the CSO spectrograph for the solar
radio spike event. In this figure, the spike signals look
like dense light-drizzle falling from the sky. The duration of
each small spike burst is about tens of milliseconds with the
minimum bandwidth being less than 10 MHz . Some spike clusters
(formed by a group of spikes) appear in the segment, and at
certain discrete frequencies the clusters have similar morphology.
For example, in the figure the arrows indicate three rows of spike
clusters with similar structures at the frequencies of $\approx$ 200 MHz,
300 MHz, and 400 MHz. There exist
many other sets of similar clusters, {\it e.g.} the ones
at $\approx$ 08:13:50 UT, at $\approx$ 270 MHz and $\approx$ 360 MHz.

\subsection{Harmonic Structures of Spike Clusters}

Four sets of spike clusters with similar structure at different
frequencies are shown in Figures 4a\ --\ b and Figures 5a\ --\ b. Figure 4a
displays a high-resolution (10 ms in time and 160 kHz in frequency)
dynamic spectrum of one radio spike cluster.
Obviously, in the figure there are two rows of spike clusters
(also called as spike chains). The frequency ranges of these two
chains are 390\ --\ 403 MHz and 285\ --\ 298 MHz,
indicated by two dashed (H4) and two dotted line pairs (H3), respectively.
Their central frequencies are 396 MHz and 291 MHz with the
frequency ratio being $\approx$ 1.36. Spike clusters within the two frequency ranges
correspond well to each other.

To quantitatively analyze the correlation between the
H3 and H4 spike cluster chains, we integrate the signal strength
of spikes within H3 and H4, and normalize the integrated intensities.
The obtained light curves are shown in the upper part of Figure 4c,
labeled by L3 for H3 and L4 for H4. The peaks of these two light
curves basically manifest a one-to-one correspondence. The
correlation coefficient CC is as high as 0.66. This indicates that
the two sets of radio spikes originate from the same energy
release and radiation process. In addition, the ratio of
central frequency of the two groups is about 1.36, very close
to 4:3, implying that the H4 and H3 spike chains are the
fourth and third harmonics of a single spike emission.

Previous studies have implicated that the occurrence of lower
harmonics has a certain time delay relative to higher harmonics
for solar radio spikes (\citealp{gudel1990}). The spike event observed
here consists of two clear cluster chains and with a high temporal
resolution, thus is suitable for further quantitative measurement of the
delay. The light curve of L3 is shifted forward or backward
by $\Delta T$ (due to the time resolution being 10 ms, $\Delta T$ is
set to be one to several 10 ms). It is found that when the light
curve of L3 is shifted forward by $\Delta T$=10 ms,
the correlation coefficient between the adjusted light curve of
L3 (referred to as L3$^\prime$) and L4 reaches the maximum value of CC$^\prime$=0.79.
This indicates
that the third harmonic emission is observed $\Delta T_{34}$ $\approx$ 10 ms
later than the fourth harmonic band. Since the temporal resolution
of the data is 10 ms, the time delay between the third harmonic
and the fourth harmonic should be in the range of 5 ms and 15 ms. Data with
a higher temporal resolution are required to further determine the
exact value of the delay.

For the radio spike in Figure 4b, we conducted a similar analysis.
The two dotted (H3) and dashed (H4) lines outline
the frequency range of 366\ --\ 390 MHz and 268\ --\ 289 MHz, respectively,
with the central frequency ratio being 1.37. In Figure 4d the correlation
coefficient CC  between the light curves of L3 and L4 is 0.69, and it becomes
as high as 0.8 after L3 being shifted forward by 10 ms.
For this row of spike chains, the time delay of the
third  relative to the fourth harmonic is also between 5 ms and 15 ms.

In the dynamic spectrum of Figure 5a, the two solid,
dotted, and dashed lines indicate three rows of radio spikes,
within the frequency range of 190\ --\ 215 MHz, 290\ --\ 330 MHz,
and 380\ --\ 430 MHz, with the ratio of the central frequency being 1:1.5:2,
corresponding to the second, third, and fourth harmonic emission,
respectively. The normalized  light curves within
the corresponding frequency band of the three groups are
demonstrated by solid, dotted, and dashed lines in Figures 5c\ --\ e, respectively.
The correlation coefficient between the second and third (the second and
fourth, and third and fourth) harmonics is 0.29 (0.25 and 0.64). When
the light curves of the second and third harmonics are shifted forward by
30 ms and 10 ms, respectively, the correlation coefficient increases to the
maximum value of 0.58 (0.8), between the second and third harmonic
bands (the third and fourth). When the light curve of
the second harmonic is moved forward by 40 ms, the correlation between the
light curves of the second and fourth harmonics is 0.54. Figure 5b shows
another group of radio spikes with the second, third, and fourth harmonics.
We also performed a similar analysis and showed the results in Table 1 together
with those of the former three groups of harmonic structures.

From Table 1, we can see the fundamental frequencies of spikes are all $\approx$ 100 MHz (not observed).
Based on the ECM mechanism, the magnetic field strength at these
spike sources is about 36 gauss with
\begin{equation}
\Omega_{\mathrm{ce}}=\frac{eB}{m_{\mathrm{e}}c}=1.759\times10^{7}B\quad(\textrm{rad}\ \mathrm{s}^{-1}),
\end{equation}
in which $e=4.8\times 10^{-10}$ ESU(C) is the electron charge,
$m_{\mathrm{e}}$=$9.1\times10^{-28}$ g is the electron mass,
$B$ represents the magnetic field strength in gauss,
and $c=3.0\times10^{10}$ cm s$^{-1}$ is the speed of light. This
provides a diagnostic of magnetic field strength in the radio source region.

\subsection{Periodic and Inverted V-shape Spike Clusters}

In Figure 4a and Figure 4b (08:11:50\ --\ 08:11:52 UT), the occurrence of
the solar radio spikes shows a nice periodicity. Such periodicity is
better seen from the light curves. There are a total of $\approx$ 10 radio peaks
within 2 s in Figure 4c and $\approx$ 15 ones within
3 s in Figure 4d (08:11:50\ --\ 08:11:53 UT). The occurrence period of
peaks is $\approx$ 0.2 s. We use wavelet analysis to analyze the two series
of light curves, and confirm the occurrence period to be 0.2 s. The confidence level
of the periodic analysis is above 95 \%.

In addition, in Figure 4a almost all of the radio spike clusters are
composed of two branches of spikes and in Figure 4c the light curves
also present a double-peak morphology. In Figure 4a the white rectangle
indicates two sets of spike clusters with clear dual-branch
structure. The radio spectrum in the white rectangular box is enlarged
and shown in Figure 6a.

Figure 6a shows that in both clusters the frequency first increases and then decreases,
forming an inverted V structure. The inverted V-shape spike clusters are found in both the
third and fourth harmonics. The start frequency of the fourth harmonic is 365 MHz, and after
50 ms, it rises to 405 MHz; 50 ms later, the frequency decreases to 387 MHz.
We can obtain that the frequency drift rate is about 800 MHz s$^{-1}$ and
- 360 MHz s$^{-1}$ for the rising and decreasing parts of the spike cluster,
respectively. These values are comparable to that of typical type III bursts.

The peak frequency of the fourth harmonic of the inverted V-shape spike
cluster is about 400 MHz, and the corresponding electron cyclotron
frequency $\Omega_{\mathrm{ce}}$ is about 100 MHz. The inferred maximum magnetic field strength
at the radio spike source is also 36 gauss.

\section{Discussion}
Two major emission mechanisms (\citealp{Benz1986}) have been proposed to explain
solar radio spikes. One is the plasma emission mechanism. This involves enhanced Langmuir
waves induced by the bump-on-tail instability caused by beams of energetic
electrons, through a complex wave-wave coupling process. These Langmuir
waves are then converted to electromagnetic radiation escaping from the source
region at the plasma oscillation frequency and its harmonics. The other one is the Electron
Cyclotron Maser (ECM) mechanism. In the magnetic structure filled with energetic
electrons injected by energy release processes in solar eruptions, some
electrons with a relatively larger parallel velocity (smaller pitch angle) may propagate
down to the solar surface and escape from the magnetic structure. This may
give rise to the electron velocity distribution function that is capable
of exciting the ECM (\citealp{wu1979}; \citealp{wu1985}; \citealp{fleishman1998};
\citealp{ergun2000}; \citealp{Bingham2013}), releasing electromagnetic waves
with frequency close to the electron cyclotron frequency of $\Omega_{\mathrm{ce}}$ or/
and its harmonics.

The observed characteristics of spike radiation (such as the narrow bandwidth, high
polarization, and high brightness temperature) can be explained by both
ECM and plasma emission  mechanisms. However, radio spike bursts can sometimes
exhibit high-order (6\ --\ 8) harmonics, which is very difficult to explain
with the plasma emission mechanism, and is generally understood as a result
of the ECM emission (\citealp{gudel1990}). Here, we use the ECM mechanism
to understand the observed time delay of the lower harmonics relative
to the higher ones of the radio
spike at metric wavelengths and the generation of
inverted V-shape spikes.

\subsection{Causes of Temporal Delay of Lower Harmonics}
There are three possible factors that can lead to the delay between
lower and higher harmonics ({\it e.g.} \citealp{dorovskyy2015};
\citealp{kong2016}). Firstly, the group speed of the lower harmonics
is smaller than that of higher ones, so it takes a longer time for the
lower harmonics to propagate to the Earth from the spike source as shown
in Figure 6b. Secondly, assuming the ECM emission arising from a central density
depleted magnetic loop  as shown in Figure 6c, higher harmonics can radiate
out from the lower site of the loop (near the emission exciting region), while
lower harmonics have to propagate to a higher site and take a longer time to
escape from the magnetic loop (like electromagnetic waves propagating in a waveguide). Last,
the refraction or reflection of electromagnetic wave at different frequencies  may also
result in the time delay of the lower harmonics.

For the first interpretation,
the group velocity $V_{\mathrm{g}}$ is determined by
\begin{equation}
V_{\mathrm{g}}=c\sqrt{1-(\frac{f_{\mathrm{pe}}}{f})^{2}},
\end{equation}
where $c=3.0\times10^{10}$ cm s$^{-1}$ is speed of light, and $f$ represents the observational
radio frequency. This frequency is taken to be $f_\mathrm{2H}$=200 MHz (the second harmonic),
$f_\mathrm{3H}$=300 MHz (third), and $f_\mathrm{4H}$=400 MHz (fourth). The frequency of $f_{\mathrm{pe}}$ is
the plasma frequency on the radio emission propagation path. Here,
\begin{equation}
f_{\mathrm {pe}}\thickapprox9000\times\sqrt{N_{\mathrm{e}}\ (\mathrm{cm}^{-3})}\quad(\mathrm{Hz}),
\end{equation}
can be deduced using the solar atmosphere electron density model of Newkirk \citep{newkirk1961} at
the distance less than 2 R$_{\odot}$:
\begin{equation}
N_{\mathrm{e}}=4.2\times10^{4}\times10^{\frac{4.32}{r}}\quad(\mathrm{cm}^{-3}),
\end{equation}
and that given by \citet{Leblanc98} for the distance range of  2\ --\ 215 R$_{\odot}$:
\begin{equation}
N_{\mathrm{e}}=3.3\times10^{5}r^{-2}+4.1\times10^{6}r^{-4}+8.0\times10^{7}r^{-6}\quad(\mathrm{cm}^{-3}),
\end{equation}
where $r$ is in unit of solar radius (R$_{\odot}$=6.955$\times10^{10}$ cm).
To remove the jump of density profiles at the connection (2 R$_{\odot}$) of the two
selected density models, we have multiplied the model of \citet{Leblanc98} by a factor of 3.8.

 The time difference of the third and fourth
 harmonics travelling to the Earth is then
 \begin{equation}
\Delta T_{34}=\frac{1}{c}\int^{215R_{\odot}}_{r_{0}}[\frac{1}{\sqrt{1-(\frac{f_{\mathrm{pe}}}
{f_{\mathrm{3H}}})^{2}}}-\frac{1}{\sqrt{1-(\frac{f_{\mathrm{pe}}}{f_{\mathrm{4H}}})^{2}}}]\mathrm{d}r,
\end{equation}
and for the second and fourth harmonics the time difference is
\begin{equation}
\Delta T_{24}=\frac{1}{c}\int^{215R_{\odot}}_{r_{0}}[\frac{1}{\sqrt{1-(\frac{f_{\mathrm{pe}}}
{f_{\mathrm{2H}}})^{2}}}-\frac{1}{\sqrt{1-(\frac{f_{\mathrm{pe}}}{f_{\mathrm{4H}}})^{2}}}]\mathrm{d}r.
\end{equation}
In order to estimate the starting height ($r_{0}$) of the emission, the following method is used.
 As we know, only radio bursts with frequency being greater than the plasma oscillation
 frequency can escape from their source region. The local plasma oscillation frequency
 can be inferred from the observed radio bursts, and then, the starting heights are
 deduced based on the solar atmosphere density models, such as, the \citet{newkirk1961} density model.
 In the event the radio spike frequencies are above 150 MHz, corresponding
 to the coronal electron density at 1.15 R$_{\odot}$. The starting heights of radio spikes
 should be higher than 1.15 R$_{\odot}$. In addition, the fundamental bands with frequency
 $\approx$ 100 MHz, corresponding to 1.25 R$_{\odot}$, are not detected, therefore, the
 emission heights are lower than 1.25 R$_{\odot}$. So the deduced
 heights of radio spikes lie in a range of 1.15\ --\ 1.25 R$_{\odot}$
 from the solar center. Putting the heights and equations 4, 5, and 6 into Equations 7 and 8, $\Delta T_{34}$
and $\Delta T_{24}$ are calculated to be  20\ --\ 10 ms and 90\ --\ 45 ms, respectively,
consistent with our observations.

It should be noted that two different coronal density models are combined to
calculate the time delay. Actually, in the range of 2\ --\ 215 R$_{\odot}$,
the calculated temporal delay is about zero, which means that it is negligible in this range.  The temporal
delay is mainly produced in the solar atmosphere within 2 R$_{\odot}$ from the solar center.

For the second interpretation, \cite{wu2002,wu2005} have used the ECM emission
within a magnetic flux tube with lower electron density in the center to
explain the emission properties of type III bursts. The latest studies have shown
that the radio spikes can be produced in large-scale magnetic loops in
active region (\citealp{cliver2011}, \citealp{morosan2016}).
Here, the cutoff
frequency of the electromagnetic wave decreases with increasing altitude. Higher
harmonics can radiate out from the density depleted magnetic loop at a
lower site (near the emission exciting region), while lower harmonics have to propagate
further out to escape and spend a longer time.

Figure 6c illustrates that the second, third, and fourth harmonics of 2 $\Omega_{\mathrm{ce}}$,
3 $\Omega_{\mathrm{ce}}$, and 4 $\Omega_{\mathrm{ce}}$  escape from the magnetic loop at
different heights. From Table 1, the fundamental frequency of the spike burst
is inferred to be about 100 MHz. We can get that the plasma frequencies at the three
heights are  200 MHz,  300 MHz, and 400 MHz (as the mode of
electromagnetic wave is unknown, we assume that the escaping frequency equals to
the O-mode cut-off frequency, {\it i.e.} the plasma oscillation
frequency). The corresponding electron densities are found to be 4.9$\times$10$^8$ cm$^{-3}$, 1.1$\times$10$^9$ cm$^{-3}$,
and 2.0$\times$10$^9$ cm$^{-3}$ at the corresponding escaping sites. These values are consistent with
the electron density of the coronal loop given by \cite{xie2017}.

In addition to the above factors, the effect of refraction and reflection cannot be
excluded. Due to the lack of information on the source of the spikes,
it is not possible to determine which cause is the most
relevant, or different causes may work together.

\subsection{Possible Cause of the Inverted V-shape Spike Clusters}
As the radio spikes appear during the flare, they should be excited by
energetic electrons accelerated via magnetic reconnection. These nonthermal electrons are
then injected into relevant magnetic structures such as magnetic loops.
When a beam of electrons moves downward along  loops, a horseshoe distribution
shall be formed with the increasing magnetic field
(\citealp{ergun2000}; \citealp{Bingham2013}). It can also be expected
that part of electrons will be reflected due to magnetic mirror
effect of magnetic loops. During the process a ring distribution of
electrons (or a loss cone) can form around the reflected
point, and then a ring-beam distribution arises when the reflected
electrons move upward (\citealp{dory1965}; \citealp{trievelpiece1968};
\citealp{wu2002}). Note that nonthermal electrons with the distributions mentioned
above are unstable and can efficiently drive ECM (\citealp{zhao2016a,zhao2016b}).
In this regard, the inverted V-shape bursts reported in the present paper are
possibly associated with electron reflection induced by magnetic
mirroring process in magnetic loops.

\section{Conclusions}
This paper reports a solar radio spike event observed by the newly
constructed high-performance metric wavelength solar radio spectrograph
at \emph{Chashan Solar Observatory}. Using the high quality observational
data, we found four sets of spike cluster chains with harmonic
structures in one solar radio spike event. The harmonic orders vary from the
second to the fourth. For the first time, the time delay of the lower
harmonics relative to the higher harmonics is determined. The time delay
is about 10 ms between the third and the fourth harmonics, and  30\ --\ 40 ms
between the second and the fourth harmonics.

Based on the ECM mechanism, the cause of the time delays is
discussed. It is found that the difference in group speeds
and propagation time in the magnetic loop of various harmonics
may cause the delay.

The coronal parameters such as the magnetic field
strength in the radio spike source region and the electron density
at the escaping sites of the radio waves are estimated. The deduced
magnetic strength is  36 gauss at the radio source.
The inferred coronal electron density at the escaping sites of the
radio waves are  4.9$\times$10$^8$ cm$^{-3}$,
1.1$\times$10$^9$ cm$^{-3}$, and 2.0$\times$10$^9$ cm$^{-3}$,
based on the assumptions that the ECM emission is from a central density
depleted magnetic loop and the escaping frequency equals to the plasma frequency.

Further observational study with higher temporal resolution (preferably with
the millisecond level) and theoretical investigation are demanded for a  more complete
understanding, so as to further reveal the diagnostic potential of solar radio spike bursts.

\begin{acks}
We thank Dr. Valentin Melnik for valuable suggestions to improve the quality of this manuscript.
The authors gratefully acknowledge the teams of RSTN and ORFEES for making their data
available to us. This work was supported by grants NNSFC-CAS U1431103, and NNSFC 41331068,
11790303 (11790300), 11503014, 41504131, 11703017,  NSF of Shandong Province (ZR201702100072, ZR2016AP13),
and was supported by the Specialized Research Fund for the State Key Laboratories.
\end{acks}

\paragraph*{\footnotesize Disclosure of Potential Conflicts of Interest} \footnotesize The
authors declare that they have no conflicts of interest.



\bibliographystyle{spr-mp-sola}
\bibliography{reference}

\IfFileExists{\jobname.bbl}{} {\typeout{}
\typeout{****************************************************}
\typeout{****************************************************}
\typeout{** Please run "bibtex \jobname" to obtain} \typeout{**
the bibliography and then re-run LaTeX} \typeout{** twice to fix
the references !}}

\newpage
\begin{table}
\caption{Parameters of four sets of solar radio spike clusters
with harmonic structures, the occurrence time, duration, central
frequency, central frequency ratio, fundamental frequency and the
time delay between different harmonic emissions are given in the
 first to the eighth columns, respectively.}
\begin{tabular}{cccccccc}
  \hline
    Time    & Duration &  Central  Freq &  Ratio   &Fund. Freq   &  \multicolumn{3}{c}{Delay (ms)} \\
     \cline{6-8}
    (UT)       &   (s) &             (MHz)      &        &    (MHz)    & $\Delta T_{23}$ &$\Delta T_{24}$  & $\Delta T_{34}$   \\
  \hline
 08:11:10     &3.5   & 390, 395       &3:4      & 97   &  $\diagdown$   &  $\diagdown$    & 10 \\
 08:11:48     & 7   & 285, 390        & 3:4     &95     & $\diagdown$  &  $\diagdown$   & 10   \\
 08:11:56     & 1   & 210, 315, 420   &2:3:4   &105    &  30           & 40            &10    \\
 08:13:34     &1    & 200, 300, 400   &2:3:4   & 100    & 20           & 30             &10   \\
  \hline
\end{tabular}
\end{table}

\newpage

 \begin{figure*}[!ht]    
   \centerline{\includegraphics[width=1.0\textwidth]{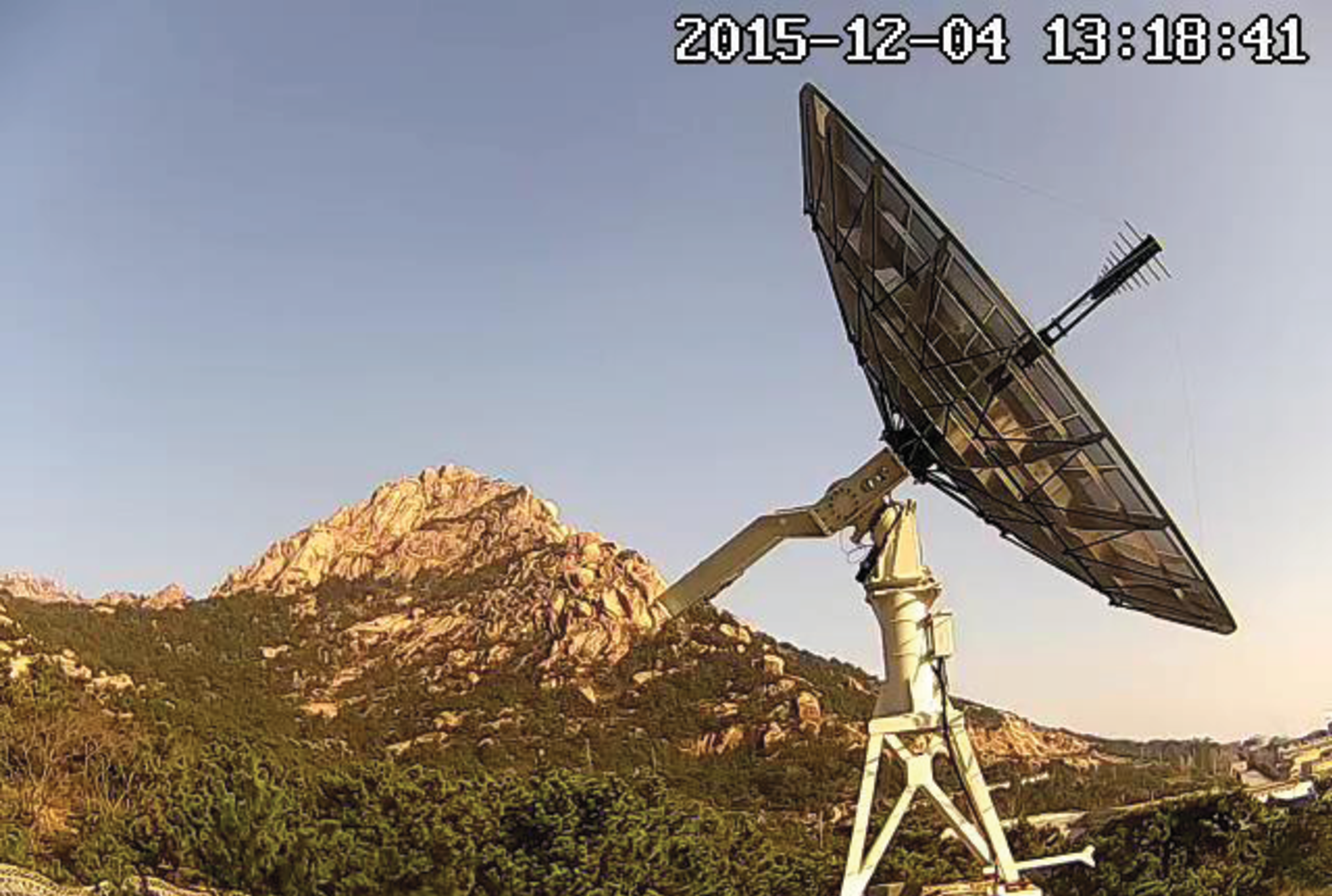}
              }
              \caption{The CSO 6-meter parabolic dish and the
              dual-polarized log-periodic antenna feed.}
   \label{f1}
   \end{figure*}
 \newpage

 \begin{figure*}[!ht]    
   \centerline{\includegraphics[width=1.0\textwidth]{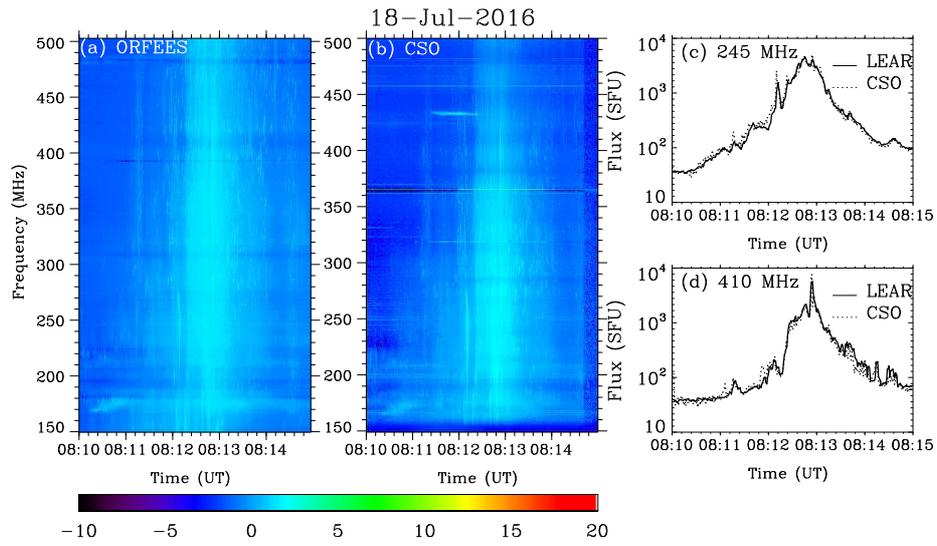}
              }
              \caption{(a, b) The
              solar radio dynamic spectra recorded by ORFEES and CSO
              at 150\ --\ 500 MHz. (c, d) Solar radio burst flux density profiles
              observed at 245 MHz and 410 MHz by the LEAR and CSO.
              See text for details. The color bar indicates the relative intensity (the same below). }
   \label{f1}
   \end{figure*}
 \newpage

\begin{figure*}[!ht]    
   \centerline{\includegraphics[width=1.0\textwidth]{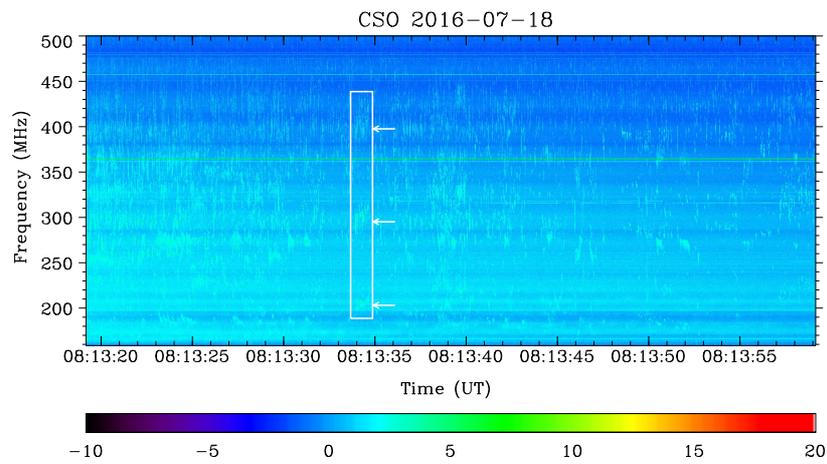}
              }
              \caption{The solar radio spikes observed by the CSO spectrograph at
              metric wavelengths. The temporal resolution of the data is 10 ms.
              The white rectangular box indicates three rows of solar radio spike
              clusters with similar structures at 200 MHz, 300 MHz,
              and 400 MHz, as pointed out by white arrows.}
   \label{f2}
   \end{figure*}
\newpage
\begin{figure*}[!ht]    
   \centerline{\includegraphics[width=1.0\textwidth]{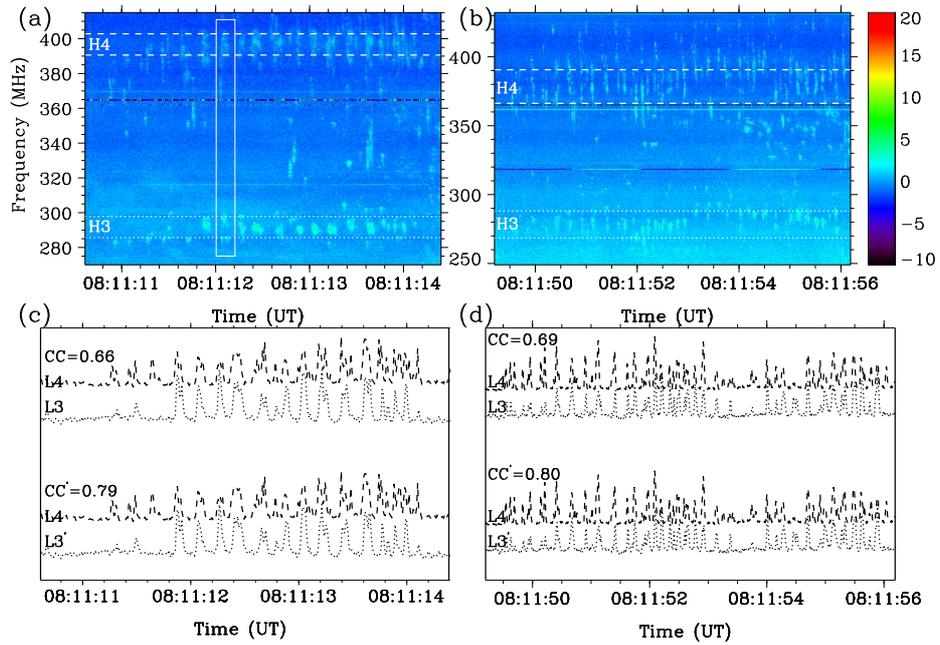}
              }
              \caption{(a, b) Spectra of two sets of solar radio spike clusters
              with the third and fourth harmonics. The dotted (H3) and dashed line
              pairs (H4) are used to indicate two spike clusters with similar
              morphology. (c) Dotted (L3) and dashed profiles (L4) are the light
              curves of integration and normalization of the radio spike burst
              intensity between the two line pairs (H3 and H4), and L3$^{\prime}$ is the
              light curve of L3 shifted forward by 10 ms. (d) Profiles are the
              same as those of (c) but obtained from the spectrum of (b).}
   \label{f3}
   \end{figure*}
\newpage
\begin{figure*}[!ht]    
   \centerline{\includegraphics[width=1.0\textwidth]{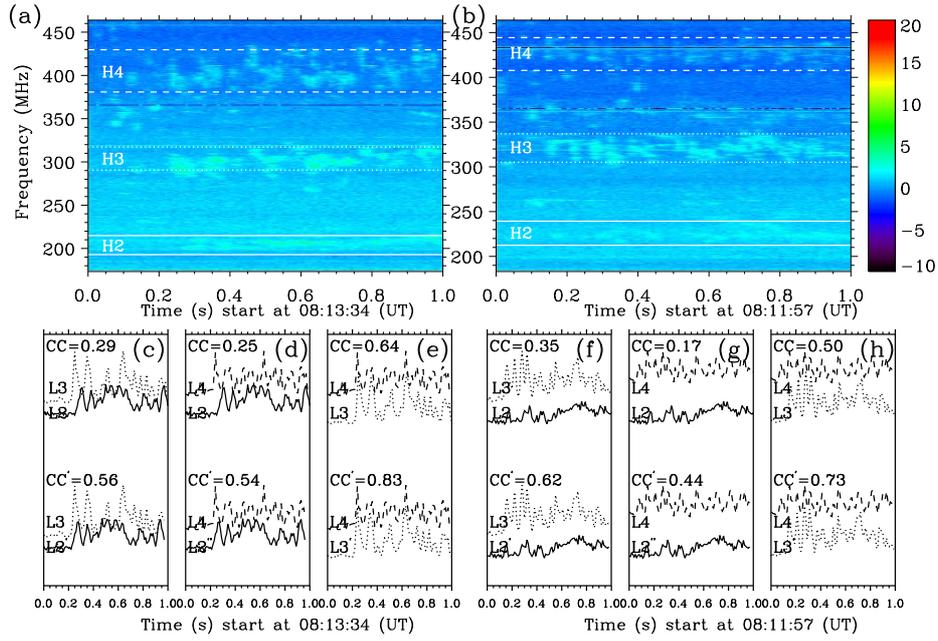}
              }
              \caption{(a, b) Spectra of two sets of solar radio spike clusters
              with the second, third, and fourth harmonics. The solid (H2), dotted (H3), and
              dashed line pairs (H4) are used to indicate three spike clusters with
              similar morphology. (c\ --\ e) solid (L2), dotted (L3),
              and dashed profiles (L4) are the light curves of integration and normalization
              of the radio spike burst intensity between the two line pairs (H2, H3, and H4),
              and L2$^{\prime}$, L2$^{\prime\prime}$, and L3$^{\prime}$ are the light curves of L2, L2, and L3 shifted forward by
              30 ms, 40 ms, and 10 ms. (f\ --\ h) Profiles
              are the same as those of (c\ --\ e) but obtained from the spectrum of (b), and
              L2$^{\prime}$, L2$^{\prime\prime}$, and L3$^{\prime}$ are the light curves of L2, L2, and L3 shifted forward by
              20 ms, 30 ms, and 10 ms.}
   \label{f4}
   \end{figure*}
\newpage
\begin{figure*}[!ht]    
   \centerline{\includegraphics[width=1.0\textwidth]{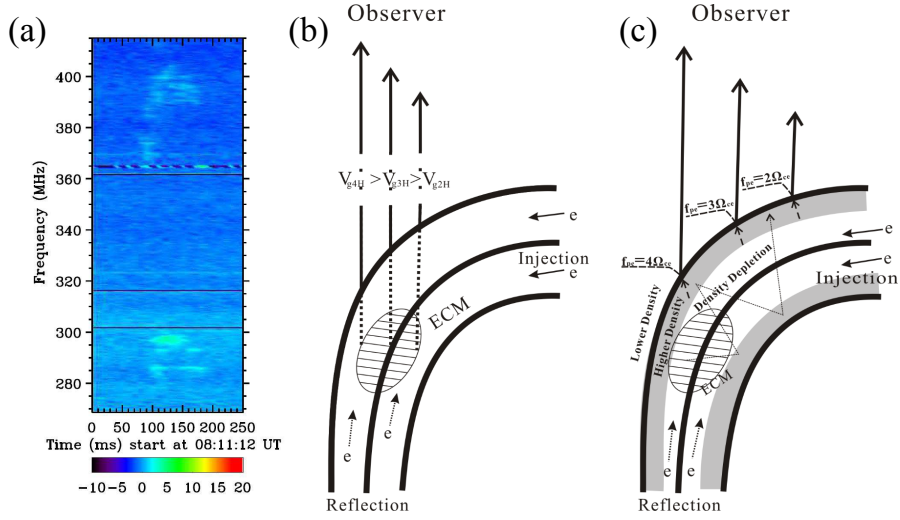}
              }
              \caption{(a) The radio spectrum of the inverted V-shape radio spike
              clusters. (b, c) The schematics for the causes of temporal delay of lower
              harmonics relative to higher orders and the formation of inverted V-shape clusters. The energetic
              electrons accelerated from magnetic field reconnection are injected
              into the magnetic loop system, and because of the magnetic mirror
              effect the electrons are reflected above the magnetic
              loop foot point. Electrons with different distributions
              are unstable and can efficiently drive ECM. The injected and reflected
              electrons result in the emission of the first and second branches of
              the inverted V-shape spike cluster, respectively. (b) The group velocities of the second, third, and fourth
              harmonics are different (V$_{\mathrm{g4H}}$ $>$ V$_{\mathrm{g3H}}$ $>$ V$_{\mathrm{g2H}}$) and this difference can
               cause the time delay. (c) When the ECM emission propagates in a centeral density depleted magnetic loop
               (the surface density is higher and the surrounding density is lower), the
               fourth harmonic escapes at the lower site with the plasma frequency $f_{\mathrm{pe}}$=4 $\Omega_{\mathrm{ce}}$,
              and the third and second harmonics escape at relatively higher sites
               of the magnetic loop structures with the plasma frequencies being
               3 $\Omega_{\mathrm{ce}}$ and 2 $\Omega_{\mathrm {ce}}$. So, the lower harmonics
               take a longer time to escape from the magnetic loop.
}
   \label{f5}
   \end{figure*}

\end{article}

\end{document}